\documentclass[twocolumn]{aastex6}
\usepackage{natbib}

\shorttitle{Orbital Stability and Precession Effects}
\shortauthors{Stephen R. Kane}

\begin{document}

\title{Orbital Stability and Precession Effects in the Kepler-89
  System}

\author{Stephen R. Kane}
\affil{Department of Earth and Planetary Sciences, University of
  California, Riverside, CA 92521, USA}
\email{skane@ucr.edu}


\begin{abstract}

Among the numerous discoveries resulting from the {\it Kepler} mission
are a plethora of compact planetary systems that provide deep insights
into planet formation theories. The architecture of such compact
systems also produces unique opportunities to study orbital dynamics
in compact environments and the subsequent evolution of orbital
parameters. One of the compact {\it Kepler} systems is Kepler-89; a
system for which the radial velocity follow-up observations place
strong upper limits on the masses of the planets and their Keplerian
orbital elements. The potential for non-circular orbits in this system
make it a compelling system to study dynamical constraints on the
measured orbital parameters. We present a dynamical analysis of the
system that demonstrates the stability of the circular model and shows
the eccentric model of the system is not stable. The analysis
indicates that planets c and d, although close to the 2:1 secular
resonance, do not permanently occupy the 2:1 resonance
configuration. We explore regions of orbital parameter space to
identify the upper bounds of orbital eccentricity for the planets. We
further show how the dynamics in the compact system leads to
significant periastron precession of the innermost planets. Finally,
we quantify the effect of the periastron precession on the transit
times of the planets compared with the cyclic variations expected from
transit timing variations.

\end{abstract}

\keywords{planets and satellites: dynamical evolution and stability --
  planetary systems -- stars: individual (Kepler-89)}


\section{Introduction}
\label{intro}

The escalation in exoplanet discoveries have revealed a large
diversity in planetary system architectures, many of which are
substantially different to the architecture of our Solar System
\citep{winn2015}. The kinds of architectures being unveiled are
largely being driven by the observational biases towards relatively
small star--planet separation, such as those biases intrinsic to the
transit method \citep{kane2008b}. For example, the discovery of many
compact planetary systems have primarily resulted from the
observations by {\it Kepler} \citep{borucki2010,borucki2016} and {\it
  K2} \citep{howell2014}. An early example of such a compact system is
that of Kepler-11, currently known to have at least six low density
planets orbiting the host star
\citep{lissauer2011a,lissauer2013}. Many of these compact planetary
systems have also been found to manifest signatures of gravitational
interactions in the form of transit timing variations (TTVs)
\citep{holman2010,holczer2016}. Such compact systems are enticing case
studies for dynamical interactions between planets and their long-term
stability \citep{funk2010,kane2015b,quarles2017,granados2018} as well
as the stability of exomoon companions \citep{kane2017c}.

The Kepler-89 (KOI-94) system is a good example of a compact system
and consists of four planets with orbital periods in the range of
3.74--54.32~days \citep{weiss2013}. The inner planet is a large
terrestrial planet with a measured density of $10.1 \pm
5.5$~g/cm$^{-3}$ whilst the outer planets are gas giants with
densities $< 1$~g/cm$^{-3}$. A combined fit of Keck/HIRES radial
velocity (RV) observations and {\it Kepler} photometry by
\citet{weiss2013}, combined with a TTV analysis by \citet{masuda2013},
yielded an orbital solution with Keplerian components. The possibility
of non-zero eccentricities introduces significantly more dynamical
interactions between the planets and the potential for regions of
instability within the orbital parameter space. Additionally, the
impact of these planetary interactions may result in precession of the
orbits that would impact subsequent transit times and durations, along
with correct interpretation of TTVs within the system.

In this paper, we present a dynamical analysis of the Kepler-89
system, that includes stability tests for a variety of Keplerian
orbital solutions and a study of the effects of periastron
precession. In Section~\ref{system} we review the architecture of the
Kepler-89 system and calculate the Hill radii and mutual Hill radii
separations for each of the planets. Section~\ref{stability} provides
the detailed results of an extended suite of stability simulations for
the system, described in terms of the short-term stability
(Section~\ref{stab1}), long-term stability (Section~\ref{stab2}), and
the periastron precession effects (Section~\ref{stab3}). The
consequences of the precession effects are described in
Section~\ref{transits}, particularly the impact on transit times and
durations in comparison to TTV amplitudes. We conclude in
Section~\ref{conclusions} and detail relevance to other compact
systems with suggestions for further work and observations.


\section{Orbital Architecture of the Kepler-89 System}
\label{system}

The stellar and planetary properties required for the study presented
here were extracted from \citet{weiss2013}, including the stellar mass
of $M_\star = 1.277 \pm 0.050$~$M_\odot$. Since the argument and times
of periastron are not provided by \citet{weiss2013}, we adopt
periastron arguments of $\omega = 90\degr$ (locations of inferior
conjunction) and periastron times equivalent to the times of
mid-transit provided by \citet{weiss2013}. Shown in
Table~\ref{planets} are the orbital period, $P$, eccentricity, $e$,
minimum planet mass, $M_p \sin i$, semi-major axis, $a$, and Hill
radius, $R_H$ for each of the known planets. The orbital periods of
planets c and d show that they are in near 2:1 resonance, which has
implications for their orbital stability. The Hill radius is given by
\begin{equation}
  R_H = r \left( \frac{M_p}{3 M_\star} \right)^{1/3}
  \label{hillradius}
\end{equation}
where $r$ is Keplerian star--planet separation
\begin{equation}
  r = \frac{a (1 - e^2)}{1 + e \cos f}
  \label{separation}
\end{equation}
where $f$ is the true anomaly. Since the Hill radius is time-dependent
for a Keplerian orbit, the Hill radius shown in Table~\ref{planets} is
the mean Hill radius (where $r = a$).

\begin{deluxetable*}{lcccc}
  \tablecolumns{5}
  \tablewidth{0pc}
  \tablecaption{\label{planets} Kepler-89 Planetary Parameters}
  \tablehead{
    \colhead{Parameter} &
    \colhead{b} &
    \colhead{c} &
    \colhead{d} &
    \colhead{e}
  }
  \startdata
  $P$ (days) & $3.743208\pm0.000015$ & $10.423648\pm0.000016$ &
  $22.3429890\pm0.0000067$ & $54.32031\pm0.00012$ \\
  $e$ & $0.25\pm0.17$ & $0.43\pm0.23$ & $0.022\pm0.038$ &
  $0.019\pm0.23$ \\
  $M_p \sin i$ ($M_\oplus$) & $10.5\pm4.6$ & $15.6^{+5.7}_{-15.6}$ &
  $106\pm11$ & $35^{+18}_{-28}$ \\
  $a$ (AU) & $0.05119\pm0.00067$ & $0.1013\pm0.0013$ &
  $0.1684\pm0.0022$ & $0.3046\pm0.0040$ \\
  $R_H$ ($10^{-3}$~AU) & 1.03 & 2.33 & 7.36 & 9.18
  \enddata
\end{deluxetable*}

\begin{figure}
  \includegraphics[angle=270,width=8.2cm]{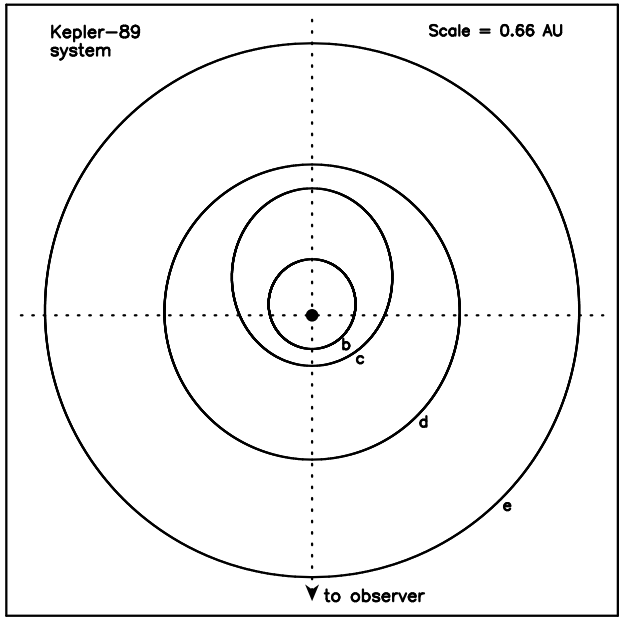}
  \caption{A top-down view of the Kepler-89 system, showing the host
    star (intersection of the dotted cross-hairs) and the orbits of
    the planets (solid lines). The Keplerian orbits are as described
    by the parameters in Table~\ref{planets} with $\omega =
    90\degr$. The scale of the figure is 0.66~AU along one edge of the
    box.}
  \label{systemfig}
\end{figure}

The orbits of the planets are shown in Figure~\ref{systemfig}. As a
first-order approximation for system stability, we use the mutual Hill
radii between adjacent planet pairs:
\begin{equation}
  R_{H,M_p} = \left[ \frac{M_{p,in} + M_{p,out}}{3 M_\star}
  \right]^{\frac{1}{3}} \frac{(a_{in} + a_{out})}{2}
    \label{mhillradius}
\end{equation}
where the ``in/out'' subscripts refer to the inner and outer planets
in the system \citep{crossfield2015,sinukoff2016}. \citet{gladman1993}
established a criterion for stability, requiring separation between
planets in two-planet systems be larger than 3.5 mutual Hill
radii. The criterion was extended by \citet{smith2009} to multi-planet
systems by defining $\Delta = (a_{out} - a_{in})/R_H$ and requiring
that $\Delta > 9$ for two adjacent planets, and $\Delta_{in} +
\Delta_{out} > 18$ for three adjacent planets where $\Delta_{in}$ and
$\Delta_{out}$ are the $\Delta$ calculations for the inner and outer
adjacent planet pairs respectively. We modify
Equation~\ref{mhillradius} to account for eccentricity with $(1 + e)$
and $(1 - e)$ multiplicative factors for $a_{in}$ and $a_{out}$
respectively. For the Keplerian orbital solution shown in
Table~\ref{planets}, we calculate $\Delta_{bc} + \Delta_{cd} +
\Delta_{de} = 3.94 + 8.97 + 11.54 = 24.45$. This indicates that the
Keplerian orbital solution is unstable as stated, particularly between
planets b and c. If we assume circular orbits for all of the planets,
we obtain $\Delta_{bc} + \Delta_{cd} + \Delta_{de} = 24.03 + 10.89 +
12.00 = 46.92$, indicating stability between pairs and overall system
stability. These stability scenarios are tested in detail in the
following section.


\section{Orbital Stability and Precession Effects}
\label{stability}

The orbital stability simulations described here were carried out
using N-body integrations with the Mercury Integrator Package
\citep{chambers1999}. We adopt a similar methodology to that used by
\citet{kane2014b,kane2016d}, which systematically explored stability
for ranges of orbital eccentricity and inclination. The dynamical
simulations made used the hybrid symplectic/Bulirsch-Stoer integrator
with a Jacobi coordinate system that generally provides more accurate
results for multi-planet systems \citep{wisdom1991,wisdom2006b} except
in cases of close encounters \citep{chambers1999}. We performed a
variety of $10^6$ and $10^7$ year integrations commencing at the
present epoch with the orbital configuration output every 100
simulation years. In accordance with the recommendations of
\citet{duncan1998}, we set the time resolution to 0.1~days to meet the
minimum required resolution of $1/20$ of the shortest orbital period
within the system. We randomize the starting positions (mean
anomalies) of the planets to create a robust snapshot of the orbital
stability for the examined architecture. Stability of simulations may
be evaluated via chaos indicators that measure the divergence of
orbits, such as the criterion developed by
\citet{cincotta1999,cincotta2000} and applied to exoplanetary orbits
\citep{gozdziewski2001a,gozdziewski2002,satyal2013,satyal2014}. Our
criteria for stability is a relatively simple first-order approach
that requires all planets survive the duration of the dynamical
simulation. This method is based on divergent orbital eccentricities
which means that the planet has either been ejected from the system or
succumbed to the host star. The simple stability criteria was compared
with chaos indicators by \citet{dvorak2010} and has been successfully
applied to exoplanetary systems in numerous instances
\citep{menou2003a,dvorak2003,kane2016d}. The additional details for
the individual simulations are outlined below.


\subsection{Short-Term Stability}
\label{stab1}

Tests for the short-term stability of the system were conducted for
$10^6$ years with both the eccentric model (Table~\ref{planets}) and
circular model ($e = 0.0$) parameters as input starting
conditions. For the eccentric model, all simulations were found to be
dramatically unstable, generally resulting in the inner (b) planet
being removed from the system within the first several hundred
years. The circular model remained stable for the full $10^6$ year
simulation in all cases and the eccentricities of the planets remained
below 0.001 for planets b, d, e, and below 0.005 for planet c.

It is worth noting that the removal of planets does not necessarily
mean that they are ejected from the system. Particularly for compact
systems such as Kepler-89, it is non-trivial for inner planets to
escape the gravitational potential well of the host star, resulting in
the planets being consumed by the host star rather than ejected.


\subsection{Near Resonance of Planets c and d}
\label{resonance}

\begin{figure*}
  \begin{center}
    \includegraphics[angle=270,width=16.0cm]{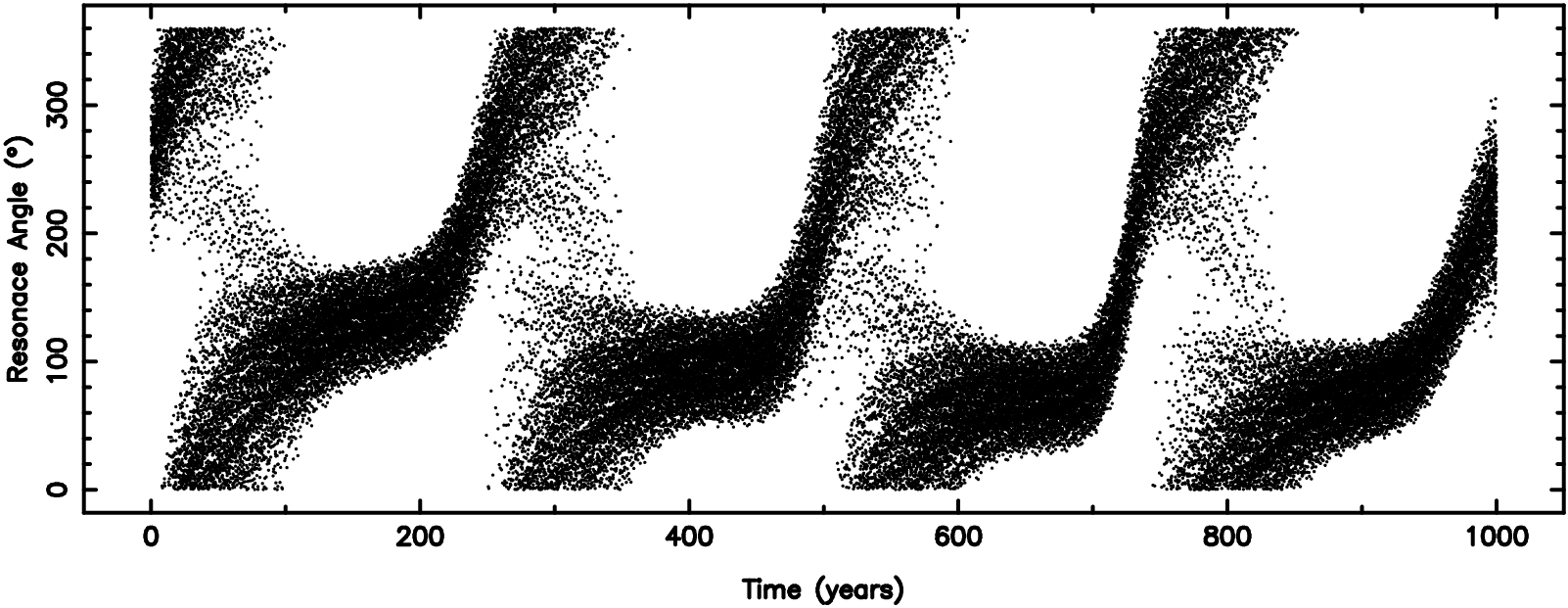} \\
    \includegraphics[angle=270,width=16.0cm]{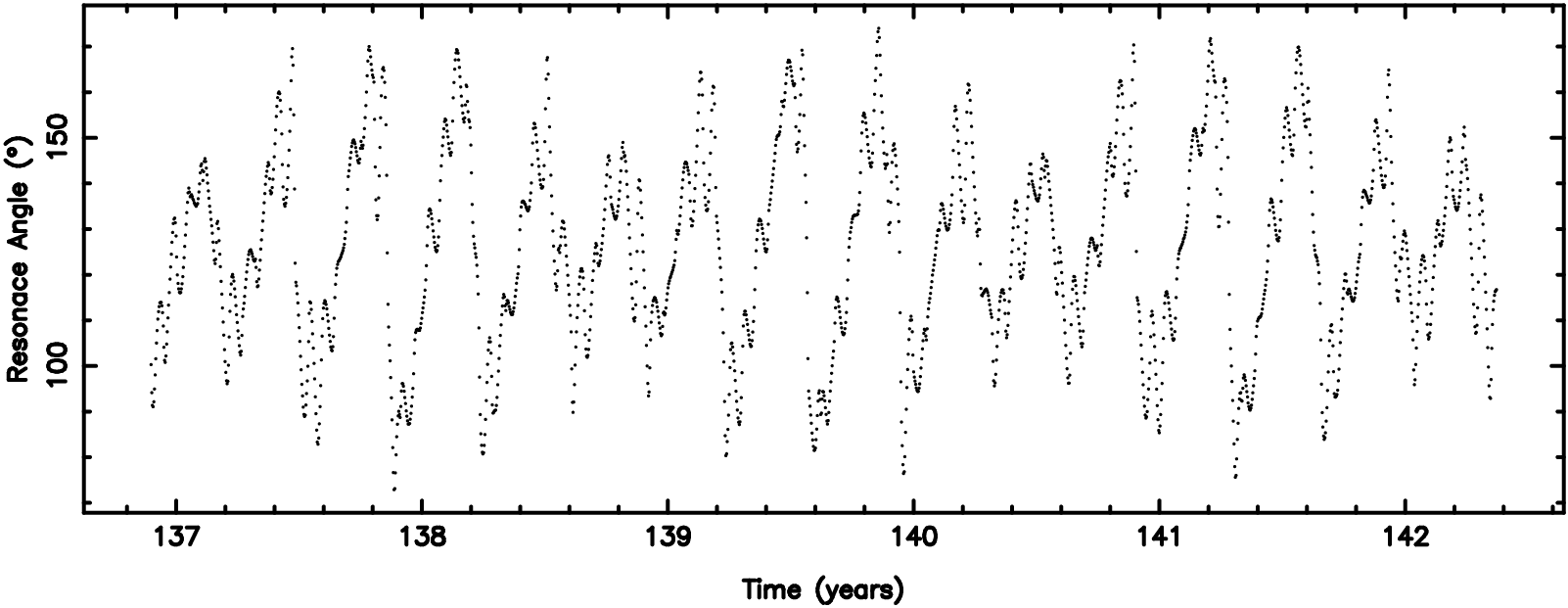}
  \end{center}
  \caption{The resonance angle for planets c and d assuming circular
    orbits, as described in Section~\ref{resonance}. Top panel: the
    results for the full $10^3$~year simulation, demonstrating that
    the planets do not remain in a stable long-term 2:1 secular
    resonant configuration. Bottom panel: a zoomed version of the top
    panel for a period when the planets are in a short term 2:1
    pseudo-resonance.}
  \label{resfig}
\end{figure*}

It was noted by \citet{weiss2013} that planets c and d are in near 2:1
resonance. We investigated the extent to which the resonance scenario
is true by executing a short-term ($10^3$ years) simulation that
assumes circular orbits for all planets, and using a high-resolution
(1~day) time output. To test for sustained resonance between the
planets, we calculated the resonance angle using the methodology of
\citet{ketchum2013} and investigated both the short-term and long-term
behavior of the c and d planetary orbits.

The results of the resonance analysis are encapsulated in
Figure~\ref{resfig}. The top panel shows the full $10^3$~year
variation of the resonance angle between planets c and d assuming a
2:1 resonance. The plot demonstrates that, although the planets are
indeed close to the 2:1 secular resonance, they regular diverge from
resonance and so do not permanently occupy the 2:1 resonance
configuration in a stable fashion. The bottom panel is a zoomed in
portion of the top panel during one of the periods of
pseudo-resonance, during which the resonance angle of the planets
oscillates with a peak-to-peak amplitude of $\sim$100$\degr$ for a
period of $\sim$80~years before the planets separate from
resonance. We additionally investigated the resonance angle of the
other planet pairs in the system but did not locate any resonant
pairs.


\subsection{Long-Term Stability}
\label{stab2}

To investigate the long-term dynamical stability of the Kepler-89
system, we extended the duration of the N-body integrations with a
variety of initial orbital architectures. The chosen duration of
$10^7$ years represents $\sim10^9$ orbital periods of the inner planet
and $\sim6.7\times10^7$ orbital periods of the outer planet. For
context, $10^7$ years represents $\sim4.2\times10^7$ orbital periods
of Mercury and $\sim6.1\times10^4$ orbital periods of Neptune. Since
it is the relative number of orbital periods that is used as a metric
for time durations relative to orbital stability, $10^7$ years is
sufficient for a compact system such as Kepler-89.

\begin{figure*}
  \begin{center}
    \includegraphics[angle=270,width=16.0cm]{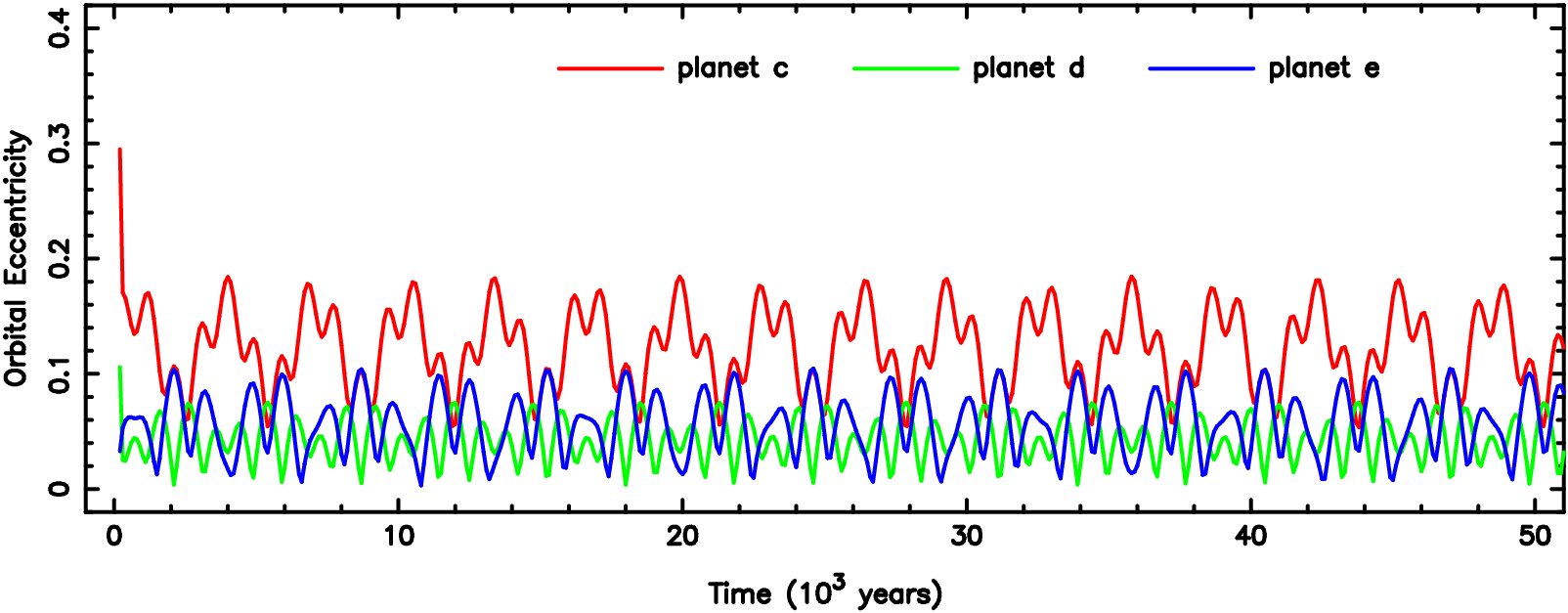} \\
    \includegraphics[angle=270,width=16.0cm]{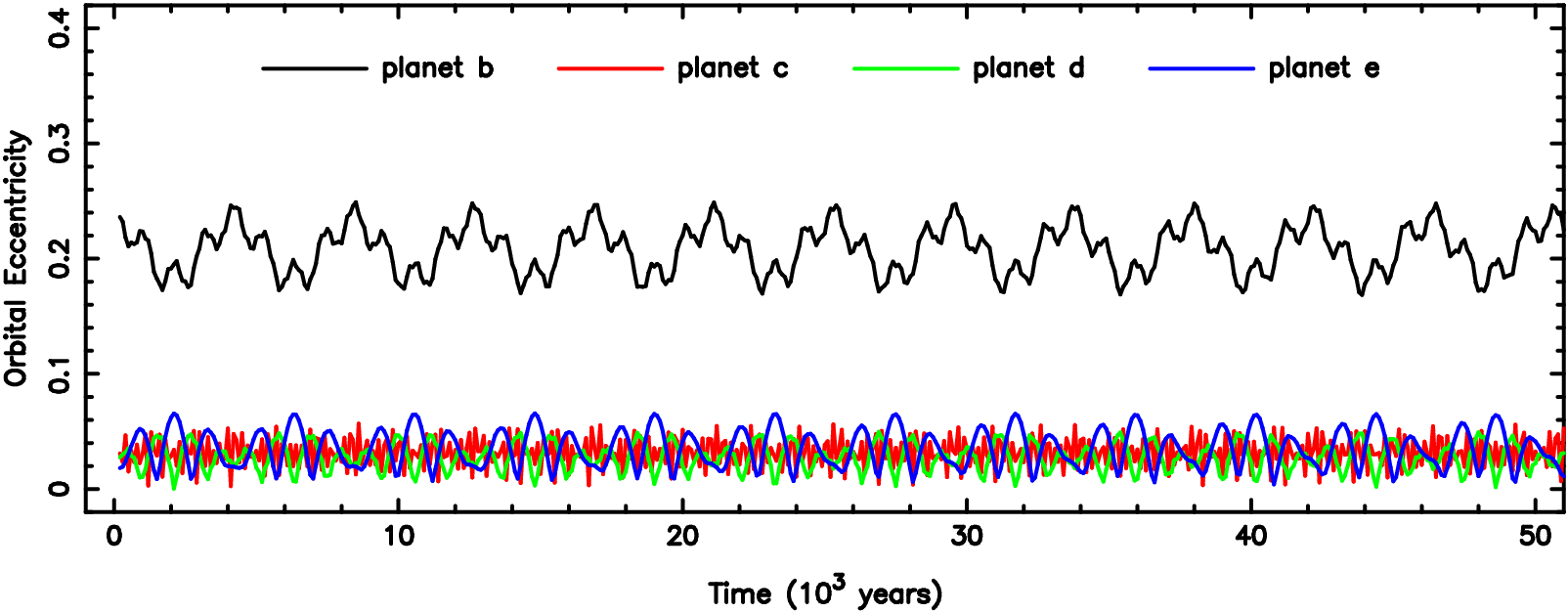}
  \end{center}
  \caption{The eccentricity as a function of time for the four planets
    of the Kepler-89 system. Top panel: assuming an eccentric model
    with planet b starting in a circular orbit. Planet b is almost
    immediately ejected from the system (after less than 200 years),
    due to perturbations resulting from interactions with planet
    c. Bottom panel: assuming an eccentric model with planet c
    starting in a circular orbit. The system is able to maintain
    long-term stability in this case.}
  \label{sim1fig}
\end{figure*}

An important aspect of the system architecture to note is that planet
c is only $\sim$50\% more massive than planet b. However, the
eccentricity of planet c is the primary cause of the system
instability for the eccentric model of the system. For example, even
if planet b begins in a circular orbit, whilst the other planets begin
with the eccentricities shown in Table~\ref{planets}, then planet b is
very quickly removed from the system due to the perturbing effects of
planet c. The outcome of this simulation is depicted in the top panel
of Figure~\ref{sim1fig}, where the three remaining planets maintain
long-term stability. The angular momentum loss from the removal of
planet b results in a dampened eccentricity of planet c which is then
periodically transferred to the two remaining outer planets. On the
other hand, if planet c begins in a circular orbit whilst all the
other planets start with their measured eccentricities then the system
is stable, shown in the bottom panel of
Figure~\ref{sim1fig}. Therefore we explored the eccentricities of
planet c that allow system stability for at least $10^7$
years. Hereafter, we refer to the eccentricities of planets b and c as
$e_b$ and $e_c$ respectively.

\begin{figure*}
  \begin{center}
    \includegraphics[angle=270,width=16.0cm]{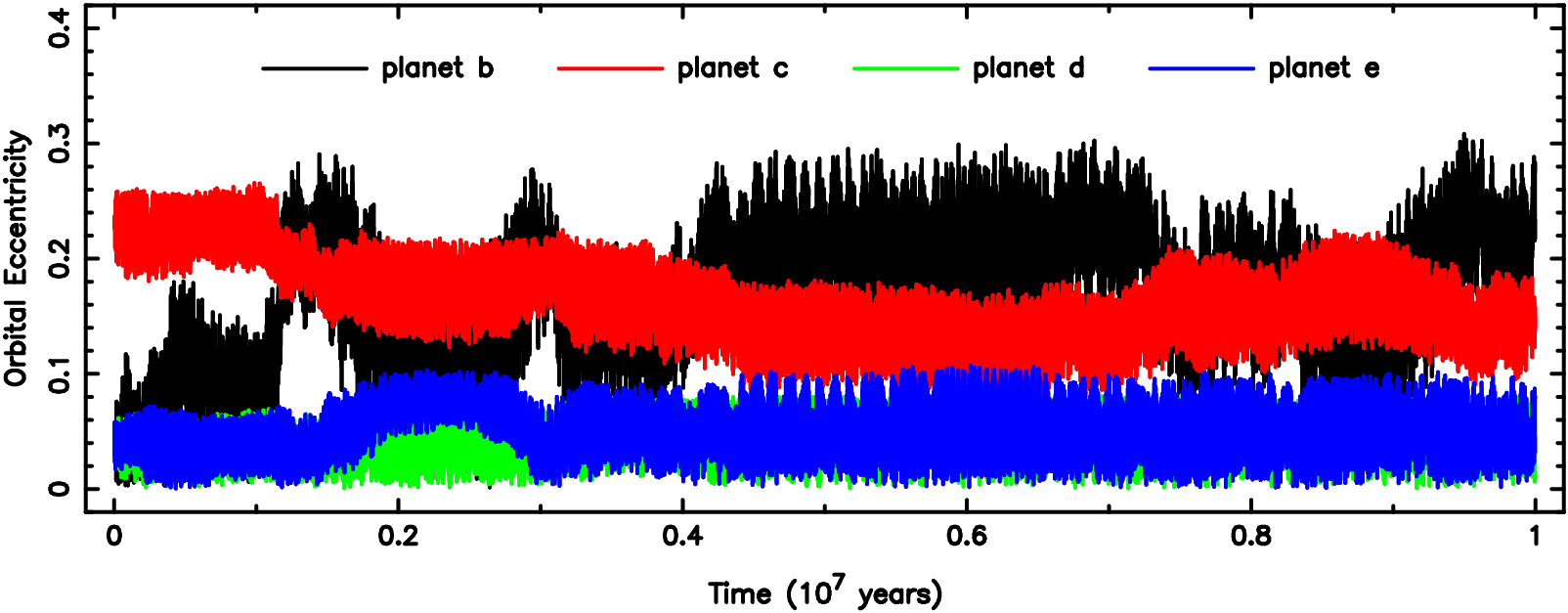} \\
    \includegraphics[angle=270,width=16.0cm]{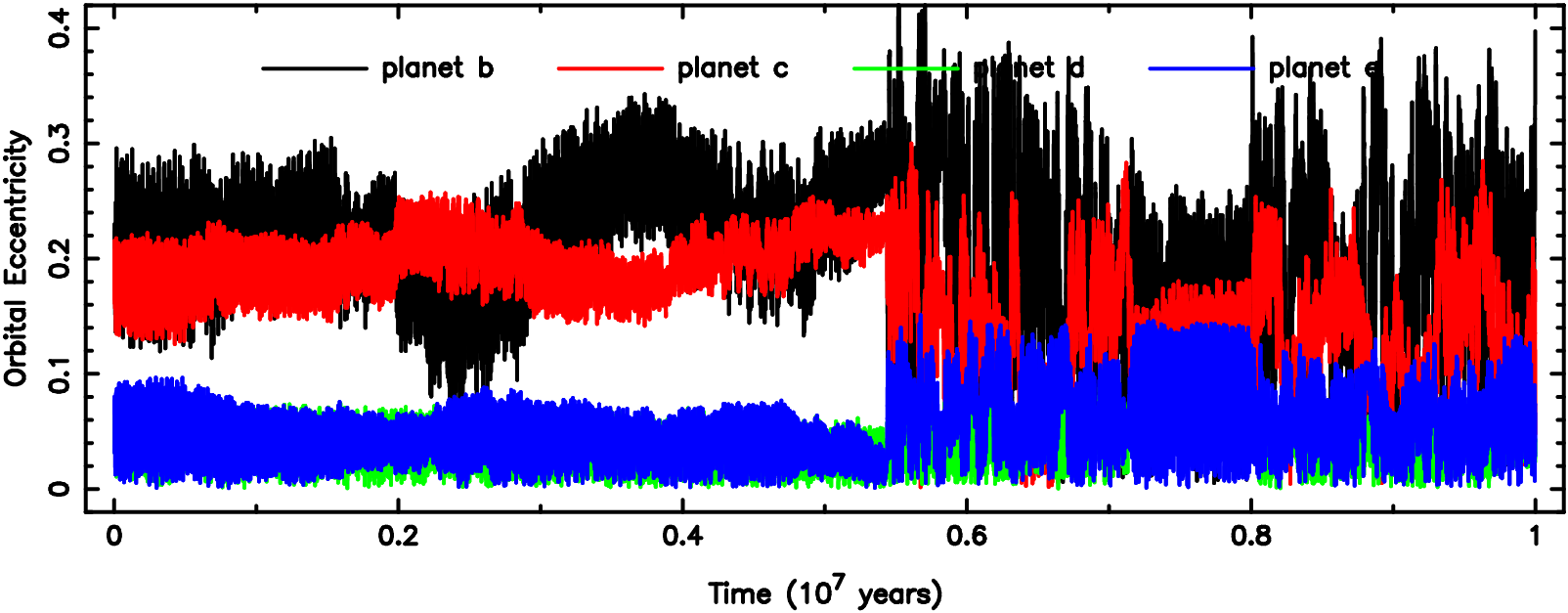}
  \end{center}
  \caption{The eccentricity as a function of time for the four planets
    of the Kepler-89 system, where the system maintained stability for
    the full $10^7$ year duration. Top panel: assuming an eccentric
    model with starting conditions for planets b and c of $e_b = 0.0$
    and $e_c = 0.26$ respectively. Planets b and c exchange
    significant amounts of angular momentum but retain orbital
    integrity. Bottom panel: assuming an eccentric model with a
    starting eccentricity of $e_c = 0.22$. This represents the maximum
    allowed starting eccentricity for planet c in the fully eccentric
    model of the system for which long-term orbital stability is
    achieved.}
  \label{sim2fig}
\end{figure*}

In the case of the eccentric model and with planet b starting a
circular orbit, the conducted simulations result in system stability
for $10^7$ years when $e_c \leq 0.26$. An example of this is shown in
the top panel of Figure~\ref{sim2fig} where $e_b = 0.0$ and $e_c =
0.26$. As can be seen, planet b cannot remain in a circular orbit due
to perturbation from the other planets. However, planets b and c
maintain a stable configuration whilst transferring angular momentum
to each other, except for a brief period at $\sim$1.5--3~Myrs where
planet e gains some angular momentum resulting in a slight increase in
eccentricity.

\begin{figure}
  \includegraphics[angle=270,width=8.2cm]{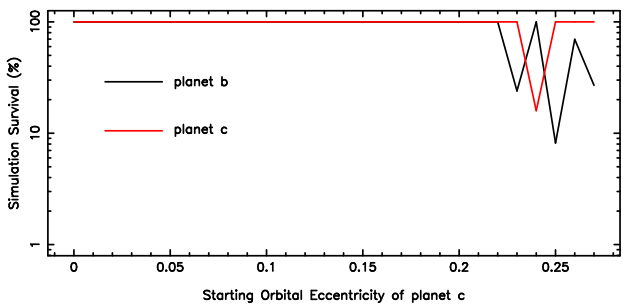}
  \caption{A summary plot of the starting eccentricities of planet c
    for the eccentric model and the resulting stability of planets b
    and c, expressed as a percentage of the total simulation time
    ($10^7$~years).}
  \label{summary}
\end{figure}

As demonstrated in Section~\ref{stab1}, the fully eccentric model of
the system is highly unstable. We gradually decreased the eccentricity
of planet c until long-term dynamical stability was achieved. We found
that the eccentric model of the system is generally stable for $10^7$
years when $e_c \leq 0.22$. An example of one of these simulations is
represented in the bottom panel of Figure~\ref{sim2fig}. It can be
seen in the figure that a relatively chaotic exchange of angular
momentum between planets b, c, and e commences after a simulation
duration of $\sim$5.5~Myrs. Planet d remains relatively unaffected by
this sudden onset of momentum exchange due to its mass being
substantially larger than the other three planets (see
Table~\ref{planets}). A further aspect worth noting is that, even
though the eccentric model of the system is unstable for $e_c > 0.22$,
not all such simulations result in the loss of planet b. In some cases
planet b survives whilst planet c is ejected. An example of this is
when $e_c = 0.24$, in which case the planetary orbital dynamics result
in the loss of planet c after $\sim$1.5~Myrs. Recall that planet c is
in a near 2:1 secular resonance with planet d (see
Section~\ref{resonance}), and the rare occasions when planet c loses
its orbital integrity is because it has been pushed out of the near
2:1 resonant island that provides temporary orbital stability
\citep{agnew2019}. Shown in Figure~\ref{summary} is a summary plot of
the variation of the starting eccentricity of planet c for the
eccentric model, from 0.0 to 0.27, and the resulting stability of
planets b and c. The stability is described as the percentage of the
simulation time ($10^7$~years) for which each planet survived. Planets
d and e are not shown in the plot since they survived in all of the
cases encapsulated by the plot.

\begin{figure}
  \includegraphics[angle=270,width=8.2cm]{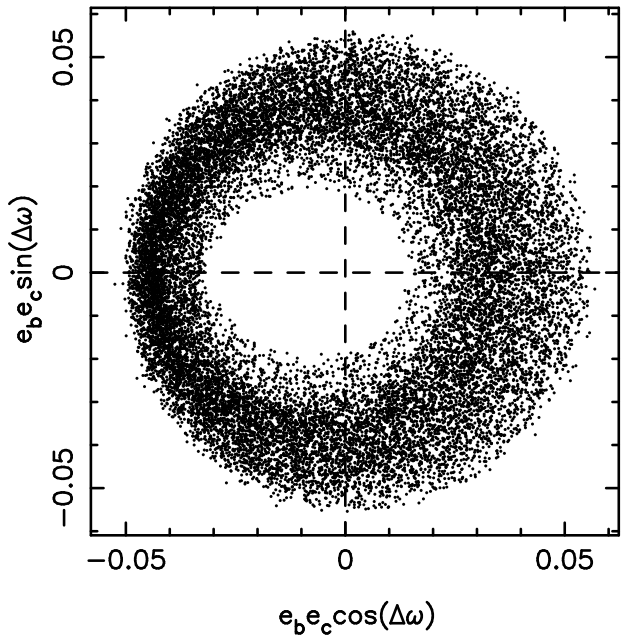}
  \caption{A polar plot of the apsidal trajectory ($e_b e_c$ versus
    $\Delta \omega$) for the b and c planets. These data represent the
    first 2~Myrs of the eccentric model with a starting planet c
    eccentricity of $e_c = 0.22$ (bottom panel of
    Figure~\ref{sim2fig}). The figure shows that the apsidal modes are
    circulating during the initial $\sim$5.5~Myrs of the simulation.}
  \label{epsilon}
\end{figure}

We further investigated the dynamical behavior of planets b and c for
the eccentric model with a starting eccentricity of $e_c = 0.22$
(Figure~\ref{sim2fig}, bottom panel) by calculating the trajectory of
the apsidal modes. Detailed descriptions of apsidal motion in the
context of interacting exoplanetary systems are provided by
\citet{barnes2006a,barnes2006c}, where the two basic types of apsidal
behavior, libration and circulation, are separated by a boundary
called a secular separatrix \citep{barnes2006c,kane2014b}. The apsidal
trajectories for planets b and c in the Kepler-89 system are
represented graphically in polar form in Figure~\ref{epsilon}. The
time span over which these are shown are the first 2~Myrs of the
simulation before the relatively chaotic transfer of angular momentum
between the system planets occurs at $\sim$5.5~Myrs. During this
stable period, Figure~\ref{epsilon} shows that the planets are
circulating since the polar trajectories consistently encompass the
origin.

It was described in Section~\ref{system} that we adopted periastron
argument values of $\omega = 90\degr$ for all planets since these are
not provided by the published orbital solutions and there is a vast
number of possible combinations. The periastron arguments were
assigned the same value to ensure that apastron for the planets occurs
on the same side of the star, thus minimizing close encounters between
the planets and optimizing the potential system stability. The
previous paragraph described how the eccentric model is only stable
where $e_c \leq 0.22$. We performed an additional test of stability by
changing the argument of periastron for the inner planet to $\omega_b
= 270\degr$. As expected, this places more stringent constraints on
the stable Keplerian orbital elements for the other planets. For this
scenario using the eccentric model described above, system stability
is only achieved for the full $10^7$ years when $e_c \leq 0.19$.

Finally, we note that the eccentricity constraints presented in this
section for the eccentric model are well within the eccentricity
uncertainties for the system, shown in Table~\ref{planets} The final
example described in the previous paragraph retains the original
eccentricities for planets b, d, and e and reduces the initial
eccentricity of planet c to $e_c = 0.22$, compared with the value of
$e_c = 0.43 \pm 0.23$ shown in Table~\ref{planets}.


\subsection{Argument of Periastron Precession}
\label{stab3}

The dynamical nature of compact planetary systems results in constant
adjustments of the Keplerian orbits for each of the planets. One of
the ways in which these adjustments are observed is through the
precession of the orbital periastron arguments. As an example of the
rate at which periastron precession can occur for the inner planets,
we conducted a $10^6$ year duration dynamical simulation with an
output interval of 1 year in order to capture the rapid pace of
periastron precession. This simulation was performed using the maximum
eccentricity orbital architecture with $e_c = 0.22$, described in
Section~\ref{stab2} and shown in the bottom panel of
Figure~\ref{sim2fig}. The precession described here includes the
effects of both general relativity (GR) and perturbations of other
planets \citep{bolmont2015}.

\begin{figure*}
  \begin{center}
    \includegraphics[angle=270,width=16.0cm]{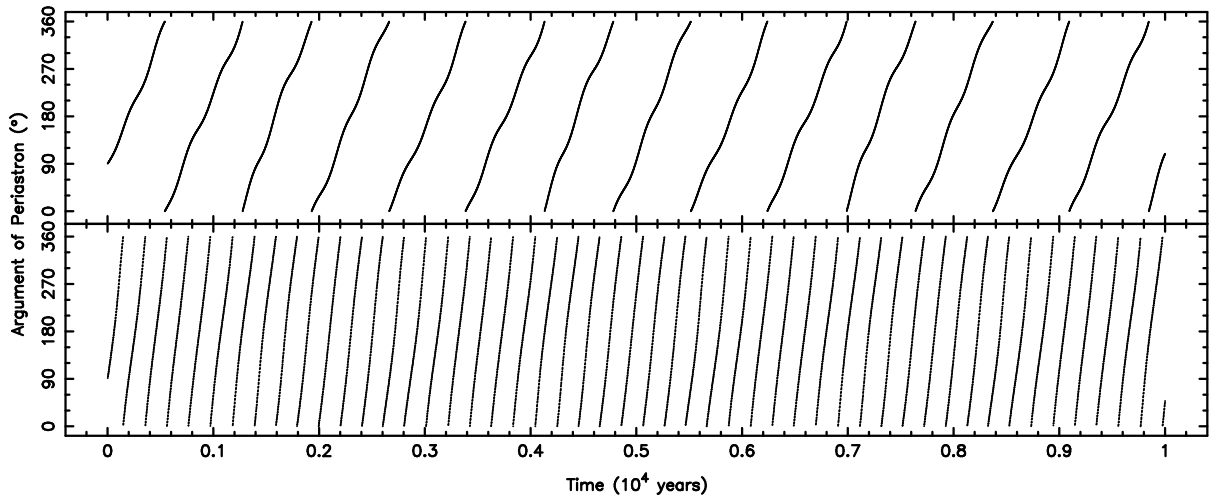}
  \end{center}
  \caption{Depiction of periastron precession for planets b (top) and
    c (bottom) for the eccentric model described in the bottom panel
    of Figure~\ref{sim2fig}. The precession rates for planets b and c
    are $0.51\degr$/year and $1.76\degr$/year respectively.}
  \label{sim3fig}
\end{figure*}

The resulting rate of change in periastron over the first $10^4$
simulation years are shown in Figure~\ref{sim3fig} for planet b (top
panel) and planet c (bottom panel). The orbits of both planets are
subjected to significant periastron precession, with planet c having a
much more rapid precession due to the combined influences of the
surrounding planets. As can be seen in Figure~\ref{sim3fig}, the rate
of periastron precession is non-uniform, but the mean rate of
precession over the $10^4$ year simulation are $0.51\degr$/year and
$1.76\degr$/year for planets b and c respectively. For comparison, the
perihelion of Mercury precesses at a rate of $0.0119\degr$/century due
to GR effects, and 0.148$\degr$/century due to perturbations from
other solar system planets \citep{clemence1947,iorio2005}. Thus the
precession rates of the planets are relatively fast, possibly
resulting in observable effects for future transit events.


\section{Effect of Orbital Dynamics on Transits}
\label{transits}

As described in Section~\ref{stab3}, the inner planets of the
Kepler-89 system undergo significant periastron precession, largely
due to planet-planet perturbations. The effect of precession on
transit times and duration has been studied by various authors
\citep{miraldaescude2002,heyl2007,jordan2008,pal2008b,ragozzine2009b,carter2010b,damiani2011,herman2018},
and the additional impact on transit probabilities has been similarly
quantified \citep{kane2012c}. Depending upon the system architecture,
the effects of the periastron precession may be observable over a
timescale less than a few years.

The Kepler-89 system has been subjected to several independent TTV
analyses \citep{masuda2013,xie2014,holczer2016,hadden2017}, including
that of the discovery paper, \citet{weiss2013}. For example,
\citet{masuda2013} found that the TTVs for Kepler-89c have a
semi-amplitude $\sim$7 minutes and a period of $\sim$155 days, caused
mostly by the aforementioned near 2:1 resonance with planet d. The
periastron precession rate of planet c described in
Section~\ref{stab3} is sufficiently large that it can impact the
calculation of the predicted transit observables. The precession rate
of planet c in the eccentric model is $1.76\degr$/year, or
$0.05\degr$/orbit. The net effect of this precession is to bring the
predicted transit time forward by $\sim$2 minutes/orbit. This effect
is generally accounted for in TTV calculations and the observability
of this effects may be utilized as an additional constraint upon
eccentricity if it is not observed in the transit data.

\begin{figure}
  \includegraphics[angle=270,width=8.2cm]{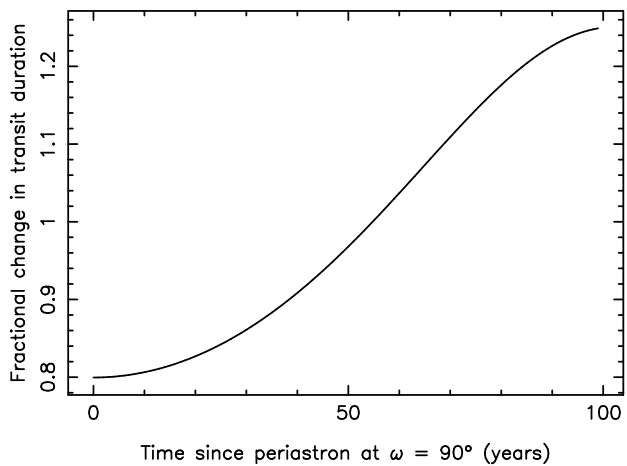}
  \caption{The fractional change in transit duration as a function of
    time (in years) since the periastron argument was located at
    $\omega = 90\degr$.}
  \label{duration}
\end{figure}

Keplerian orbital elements can also have a significant effect on the
duration of a planetary transit \citep{barnes2007d,burke2008a}. From
\citet{burke2008a}, the duration of a transit scales according to the
following relation:
\begin{equation}
  \tau = \frac{\sqrt{1-e^2}}{1+e\cos(\omega-90\degr)}
\end{equation}
Combining this scaling relationship with periastron precession results
in a time-dependent transit duration. For Kepler-89c, the rate of
precession results in $\sim$$180\degr$ change in the argument of
periastron over the course of 100 years. This is represented in
Figure~\ref{duration}, where the fractional change in the transit
duration is plotted against the time (in years) since the periastron
argument was location at $90\degr$. For the eccentricity of $e_c =
0.22$, the transit duration has a fractional change of $\sim$0.4
between inferior and superior conjunction, or a fractional change of
0.004/year 1 part in 10,000 per orbit. Thus, although the effect of
periastron precession on transit duration is unlikely to be detected
between orbits, the effect could be reasonably observed with further
transit observations in subsequent years.


\section{Conclusions}
\label{conclusions}

The orbital dynamics of compact planetary systems is a topic of
crucial importance at the present time, since current detection
methods are biased towards the detection of such systems and the
extraction of reliable masses via TTVs is dependent upon the mutual
interactions of the planets. Orbital stability in compact systems is
exceptionally sensitive to the eccentricities of the individual
planets, which in turn are difficult to measure for the majority of
{\it Kepler} systems due to the relative faintness of the host
star. The Kepler-89 system is an example of this, where the Keplerian
orbital parameters are largely determined from the {\it Kepler}
photometry rather than the RV data, whose utility is mostly to provide
a mass estimate for the giant planet (planet d).

In this work, we have used the Kepler-89 system as a template from
which it is demonstrated how the orbital eccentricity may be
constrained through the use of dynamical simulations. The results
presented here demonstrate that dynamical simulations may be used as a
powerful tool to explore numerous possible dynamical architectures
that result from systems. For example, we show that there is a limited
range of eccentricities for planets b and c (e.g., $e_c \leq 0.22$ for
the fully eccentric model) that ensure long-term dynamical stability,
and we have shown how the eccentricities vary with time for several of
the stable orbital architectures. We have further demonstrated the
dramatic periastron precession that may be occurring within the
Kepler-89 system; a result of both the compact architecture combined
with eccentricities that lie at the upper boundaries of stability. The
periastron precession for Kepler-89c is significant enough that it
should result in observable effects, or else rule out the eccentricity
that would produce such precession. In either case, the precession
effects should be careful considered when performing accurate transit
timing and duration measurements.

The near 2:1 secular resonance of planets c and d is an important
aspect of the overall system architecture, and particularly important
since planet d is the dominant planetary mass in the system. As shown
in Section~\ref{resonance}, planets c and d are not quite in resonance
but occasionally exhibit resonant behavior.  A 2:1 secular resonance
would raise the question of whether the system could dynamically
harbor a three-body Laplace resonance with a 4:2:1 orbital period
ratio. There are several known examples of Laplace resonances among
exoplanetary systems, such as GJ~876 \citep{rivera2010b}, Kepler-60
\citep{gozdziewski2016}, and TRAPPIST-1 \citep{luger2017b}. It was
also proposed by \citet{libert2013a} that Laplace resonances could be
detected via long-term observations of TTV effects. In order for
Kepler-89 to contain such a resonance that extends the existing near
2:1 secular resonance of planets c and d to a Laplace configuration,
an additional planet would need to be located close to either a 5~day
or 44~day orbital period. Given the relatively tenuous stability of
the b and c planets and the high mass of the d and e planets, such a
Laplace resonance may be difficult to achieve, but nonetheless
important for future observers of the system to keep in mind.

Many uncertainties remain in the Keplerian orbital elements of the
Kepler-89 system that will require further RV observations and/or
precise transit times to resolve. However, the compact nature of the
system combined with the possibility of non-zero eccentricities make
the system a good case-study for exploring the dynamical effects of
the orbits on each other and observable effects. The dynamical effects
include gravitational perturbations, precession of orbits, mean motion
resonances, and secular resonances. Strategies for the refinement of
planetary orbits are being applies to numerous known exoplanetary
systems \citep{kane2009c}, and could be further applied to systems
that are predicted to produce similar observable signatures. It is
possible that the time baseline of {\it Kepler} observations is too
short to observe the discussed precession effects, but observations of
the {\it Kepler} field by the Transiting Exoplanet Survey Satellite
({\it TESS}) may sufficiently extend the baseline to reveal
eccentricities hidden in the photometry.


\section*{Acknowledgements}

The author would like to thank Lauren Weiss for several insightful
discussions regarding the Kepler-89 system, and also Rory Barnes, Paul
Dalba, Sean Raymond, and Paul Wiegert for useful feedback on the
manuscript. Thanks are also due to the anonymous referee, whose
comments greatly improved the quality of the paper. This research has
made use of the following archives: the Habitable Zone Gallery at
hzgallery.org and the NASA Exoplanet Archive, which is operated by the
California Institute of Technology, under contract with the National
Aeronautics and Space Administration under the Exoplanet Exploration
Program. The results reported herein benefited from collaborations
and/or information exchange within NASA's Nexus for Exoplanet System
Science (NExSS) research coordination network sponsored by NASA's
Science Mission Directorate.


\software{Mercury \citep{chambers1999}}




\end{document}